\documentclass[fleqn,usenatbib]{mnras}
\usepackage[dvipsnames]{xcolor}

\usepackage[T1]{fontenc}
\usepackage{graphicx, amsmath, amssymb, ulem}

\newcommand{\E}{{\cal E}}

\renewcommand\textbf[1]{\textcolor{black}{#1}}
\newcommand\Msun{{\rm M_\odot}}

\title[Constraining Merging Galaxy Clusters]{Constraining Merging Galaxy Clusters with X-ray and Lensing Simulations \& Observations: The case of Abell 2146}

\author[U. Chadayammuri et al.]{
Urmila Chadayammuri,$^{1, 2}$\thanks{E-mail: urmila.chadayammuri@cfa.harvard.edu}
John ZuHone,$^{1}$
Paul Nulsen,$^{1}$
Daisuke Nagai,$^{2,3}$\newauthor
Sharon Felix,$^{4}$
Felipe Andrade-Santos,$^{1}$ 
Lindsay King,$^{4}$
Helen Russell$^{5}$
\\
$^{1}$Chandra X-ray centre, Smithsonian Astrophysical Observatory, 60 Garden Street, Cambridge, MA 02143, USA\\
$^{2}$Department of Astronomy, Yale University, New Haven, CT 06511, USA\\
$^{3}$Department of Physics, Yale University, New Haven, CT 06520, USA\\
$^{4}$Department of Physics, University of Texas - Dallas, Richardson, TX 75080, USA\\
$^{5}$School of Physics and Astronomy, University of Nottingham, Nottingham NG7 2RD, UK
}

\date{Accepted XXX. Received YYY; in original form ZZZ}

\pubyear{2021}

\begin{document}
\label{firstpage}
\pagerange{\pageref{firstpage}--\pageref{lastpage}}
\maketitle

\begin{abstract}
Galaxy cluster mergers are a powerful laboratory for testing cosmological and astrophysical models. However, interpreting individual merging clusters depends crucially on their merger configuration, defined by the masses, velocities, impact parameters, and orientation of the merger axis with respect to the plane of the sky. In this work, we investigate the impact of merger parameters on the X-ray emitting intracluster medium and gravitational lensing maps using a suite of idealised simulations of binary cluster mergers performed using the GAMER-2 code. As a test case, we focus on modeling the Bullet Cluster-like merging system Abell 2146, in which deep \textit{Chandra} X-ray and lensing observations revealed prominent merger shocks as well as the mass distribution and substructures associated with this merging cluster. We identify the most interesting parameter combinations, and evaluate the effects of various parameters on the properties of merger shocks observed by deep \textit{Chandra} and lensing observations. We show that due gravitational compression of the cluster halos during the merger, previous mass estimates from weak lensing are too high. The plane of the merger is tilted further from the plane of the sky than estimated previously, up to $30^\circ$ from the plane of the sky. We discuss the applicability of our results to multi-wavelength observations of merging galaxy clusters and their use as probes of cosmology and plasma physics.
\end{abstract}


\section{Introduction}
Merging galaxy clusters are powerful astrophysical laboratories for studying cosmology and astrophysics. To date, merging galaxy clusters have provided unique constraints on the nature of dark matter \citep{Markevitch2004,Clowe2006,Massey2015,Harvey2015,Massey2018} and on the plasma physics of the X-ray emitting intracluster medium \citep[ICM; see][for reviews]{Markevitch2007,ZuHone2016}. Mergers are crucial to the hierarchical formation of galaxy clusters, and the frequency of mergers with different mass ratios depends on cosmology \citep{Lacey1993,Fakhouri2010}.
Spatial offsets between the collisionless stars, dissipative gas, and lensing mass in merging systems like the Bullet Cluster provide constraints on the cross-section of self-interaction of the dark matter \citep[][for a review]{Randall2008,Kahlhoefer2014,Kim2017,Robertson2017,Tulin2018}. Shocks and cold fronts are also used to derive unique constraints on the microphysics of the ICM, such as the rate of electron-proton equilibration \citep{Rudd2009,Avestruz2015,Wang2018}, thermal conduction \citep{Markevitch2003,ZuHone2013}, and viscosity \citep{Roediger2013,ZuHone2015,Schmidt2017}; each of these in turn constrains the properties of cosmic magnetic fields \citep[see e.g.,][for a review]{Brunetti2014}.

Cosmological simulations yield distributions for the merger speed and dark matter concentrations of the halos \citep{Neto2007, Duffy2008}, which in turn \textbf{provide priors for} dynamical parameters for merging clusters. However, interpreting specific observed features associated with merging clusters is particularly challenging, because (a) the initial merger and structural parameters are unknown and (b) the systems are not in equilibrium \citep{Golovich2016, Golovich2017, Wittman2018}. Understanding and controlling the effects of these parameters are crucial for using merging galaxy clusters as probes of cosmology and plasma physics.

Idealised simulations enable powerful, controlled experiments to explore the large space spanned by the cluster merger parameters as well as non-gravitational processes operating during the cluster formation and evolution \citep{Ricker2001, Ritchie2002, Poole2006, ZuHone2011b}. For example, the halo masses, infall velocity and impact parameter largely determine the curvatures and Mach numbers of the shocks. The observed features also depend critically on the direction from which the merging cluster is viewed.
Pre-shock gas temperatures depend on the masses of the two substructures, as well as their initial gas profiles. The strength of a merger shock depends on the velocity of the perturbing subcluster core through the ICM of the primary cluster, and on the time, or merger phase, at which the system is observed \citep{Zhang2019, Zhang2020}. Furthermore, the observed strength of a shock and the curvature of the shock front both decrease as our viewing direction deviates from the normal to the plane of the merger. Due to the large number of parameters and potential degeneracies between them, they cannot be determined analytically. It is necessary to explore the multi-dimensional parameter space with tailored simulations.

In this work, we use simulations to understand one of the best-observed merging galaxy clusters in the X-ray. The ICM of Abell 2146 was first observed with the \textit{Chandra} X-ray Telescope in 2009 \citep{Russell2010}. This observation revealed some of the clearest merger shocks since the Bullet Cluster \citep{Markevitch2002}. Being less massive and thus cooler than the Bullet Cluster, the gas in Abell 2146 radiates in an energy range where the effective area of \textit{Chandra} is higher, so that surface brightness and temperature can be mapped in unprecedented detail. Constraints on the mass ratio $R = M_2:M_1$ of the system have been made in previous work using weak lensing and galaxy velocities \citep{King2016, White2015}, one projection angle was inferred from the line-of-sight velocity separation between galaxies in different subclusters, and infall velocities and time since pericentre passage have been estimated from the positions and Mach numbers of the two shocks in the X-ray \citep{Russell2012}. 

In order to determine the parameters of the merger in Abell 2146, we perform a suite of idealised simulations of binary cluster mergers using the GPU-accelerated adaptive mesh refinement code GAMER-2 \citep{Schive2018}. We use the most quickly evolving observables to narrow down the times at which simulated mergers best match Abell 2146, and then assess the effects of the remaining parameters and viewing direction on the observable X-ray and lensing properties of the simulated clusters.

We describe the simulation setup and translation to observable quantities in \S\ref{sec:sims}. We present observational constraints from X-ray and lensing observations of A2146 in \S\ref{sec:obs}. We describe the parameter tests in \S\ref{sec:results}, concluding with best fit parameters for Abell 2146 and summarize conclusions in \S\ref{sec:conclusions}.
Throughout this paper we use the flat \citet{Planck2016} cosmology with $H_0 = 67.7\rm\ km\,s^{-1}\,Mpc^{-1}$ and $\Omega_{\rm m} = 0.307$. At the redshift of Abell 2146, $z = 0.2323$, this corresponds to an angular scale of $3.823\rm\  kpc/arcsec$.

\section{Simulations}

\label{sec:sims}

The simulations were run with GAMER-2 \citep{Schive2018}, a GPU-accelerated Adaptive MEsh Refinement code. \textbf{The GPU acceleration allows us to simultaneously explore the effect of an unprecedented number of distinct parameters. For this initial study, we use $\sim$ 25 million dark matter particles, and use four levels of refinement to achieve a hydrodynamic resolution of 6.8~kpc.} Each run takes 5 hours on the Nvidia Tesla K80 GPU nodes on the Grace cluster at the Yale High Performance Computing centre. \textbf{The present suite of simulations models the dark matter and non-radiative gas. }Here, we describe the initial conditions and set up of the merger simulations.

\subsection{Dark matter and gas profiles}
We model the initial dark matter distribution of a cluster with the ``super-NFW'' profile:
\begin{align}
    \rho(r) &= \frac{3M}{16\pi a^3}\frac{1}{(r/a)\times(1+r/a)^{5/2}}
\end{align}
where the scale radius is related to the half-mass radius as $a = R_e/5.478$ \citep{Lilley2018}. This form has a finite total mass,\textbf{ which makes it preferable to the NFW profile for ensuring a smooth cutoff of the DM density at large radii. sNFW properties are related to those of the more widely used NFW profile as follows. The scale radius $r_{sc}$ of a halo is defined as the radius where the logarithmic slope of the density profile is $d\log\rho/d\log r=-2$,
and the concentration is defined as $c_{\rm NFW} = r_v/r_{sc}$, where $r_v$ the virial radius. For NFW halos, \citet{Duffy2008} measured the concentration-mass relation}:
\begin{align}
    c_{\rm NFW}= 5.74 \times \left(\frac{M}{2 \times 10^{12}h^{-1} M_\odot}\right)^{-0.097},
\end{align}
\textbf{where the distribution in $c_{\rm NFW}$ at fixed $M$ is log-normal and the standard deviation of $\log_{10} c_{\rm NFW}$ is 0.14. For the sNFW profile, \textbf{$r_{sc} = 2a/3$} so that the concentration is $c_{sNFW} = r_v/r_s = 3r_v/2a$. Fitting sNFW profiles with NFW formulae, \citet{Lilley2018} find that the concentrations in the two models are well-described by a linear fit, $c_{sNFW} = 1.36 + 0.76 c_{\rm NFW}$. Therefore, given an NFW concentration, we find the equivalent sNFW concentration, which then yields the scale radius $a$.}

The gas is initially set up to be in hydrostatic equilibrium with the dark matter. This criterion alone, however, is insufficient to yield both temperature and density profiles. We therefore model the gas density with the modified beta profile of \citet{Vikhlinin2006}, which, along with the condition of hydrostatic equilibrium, gives the temperature. These profiles have the form
 \begin{align}
 n_pn_e &= n_0^2\frac{\left(r/r_c\right)^{-\alpha}}{\left(1+r^2/r_c^2\right)^{3\beta - \alpha/2}} \frac{1}{\left(1+r^\gamma/r_s^\gamma\right)^{\epsilon/\gamma}}
 \label{eq:vikhlinin}
 \end{align}
where $n_e$ and $n_p$ are the number densities of electrons and protons, respectively. The inner density slope $\alpha$ and the core radius $r_c$ control the strength and extent of the cool core. The scale radius of the gas density, $r_s$, is independent from the sNFW scale radius $a$. At intermediate radii the slope of the density profile, in log-log space, is - 3$\beta$, and outside $r_s$ it transitions to -(3$\beta$ + $\epsilon$/2) over a length scale determined by $\gamma$. \textbf{Once the total mass profile is set, the dark matter particles are given velocities that place them in virial equilibrium, using the procedure outlined in \citet{Kazantzidis2004}, where the energy
distribution function is calculated via the Eddington formula \citep{Eddington1916}:
\begin{equation}
{\cal F}(\E) = \frac{1}{\sqrt{8}\pi^2}\left[\int^\E_0{d^2\rho \over d\Psi^2}{d\Psi \over \sqrt{\E - \Psi}} + \frac{1}{\sqrt{\E}}\left({d\rho \over d\Psi}\right)_{\Psi=0} \right]
\end{equation}
where $\Psi = -\Phi$ is the relative potential and $\E = \Psi - \frac{1}{2}v^2$
is the relative energy of the particle. We compute this distribution function and use it to determine DM particle speeds using the acceptance-rejection method. The direction of each particle's velocity is determined by choosing random unit vectors in $\Re^3$. The gas cell velocities are zero, i.e., the system is in hydrostatic equilibrium.}

In \S \ref{sec:gasprof}, we explain how $r_c$ was chosen to roughly match observations of relaxed cluster profiles. Since this study focuses on merger shocks, we do not vary the model parameters $r_s/a$, $\gamma$ or $\epsilon$, which affect the gas distribution on large scales. These would likely be relevant if the merger were more evolved and the shocks much further out, or in modeling a system with accretion shocks. Both these scenarios are outside the scope of the current work. We adopt $\beta$ = 2/3, $\gamma$ = 3 and $\epsilon$ = 3, which were found to fit all the observed clusters in the \citet{Vikhlinin2006} sample, and set $r_s=a$. 

\subsection{Merger geometry}
\label{sec:geom}
\begin{figure}
    \centering
    \includegraphics[width=0.5\textwidth]{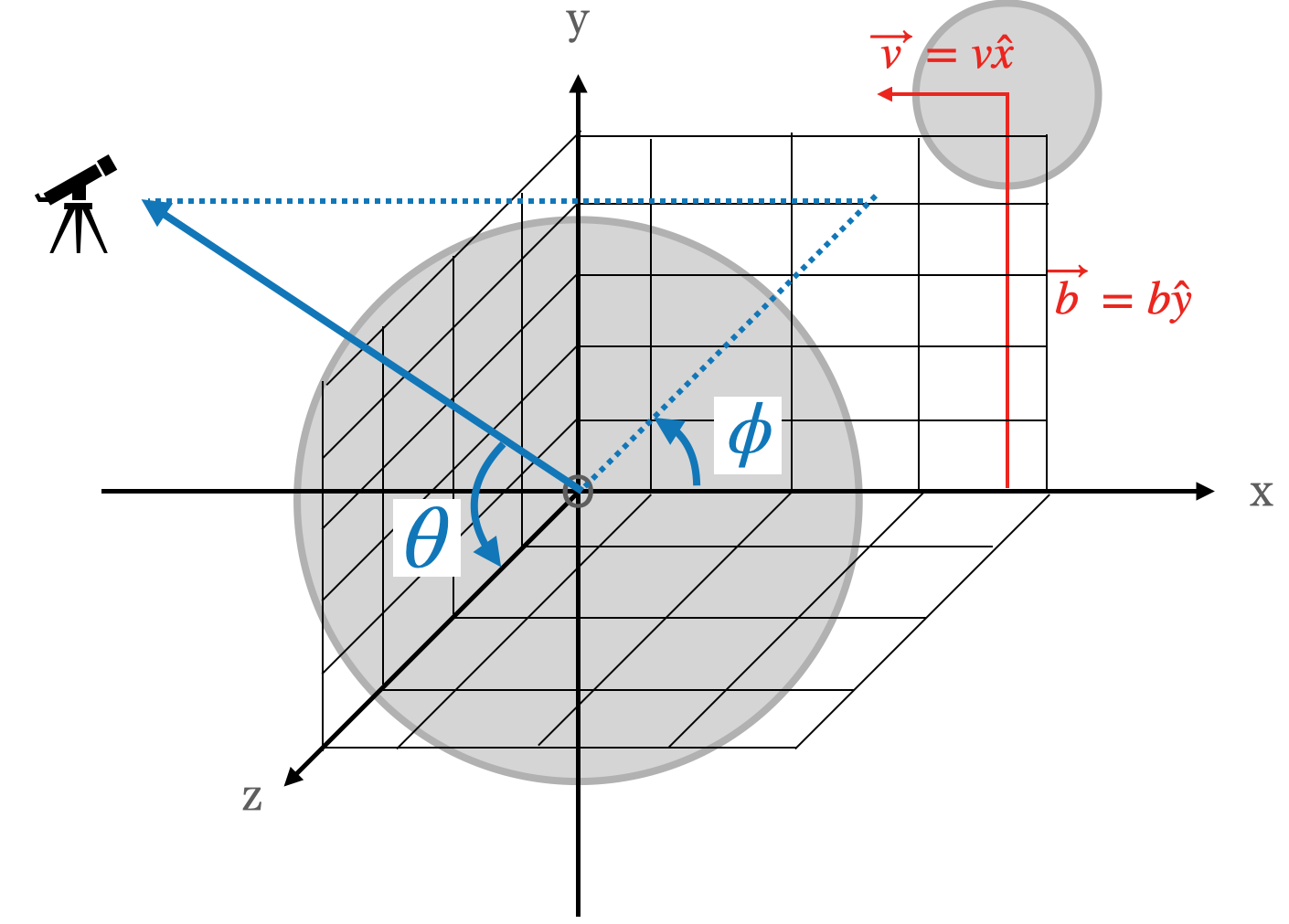}
    \caption{Schematic of initial configuration. The relative velocity between the subclusters is in the \textrm{x} direction, with an initial impact parameter $b$ along the \textrm{y} direction; the merger thus occurs in the $x-y$ plane. The observer is in the direction of the telescope, so that the viewing direction is determined by the polar angle $\theta$  and the azimuthal angle $\phi$.}
    \label{fig:geometry}
\end{figure}

The merger evolution and observed properties depend on the impact parameter $b$ and relative velocity $v$---their product $L$ is the specific angular momentum of the subhalo. Initially,the primary halo sits on the $x$-axis and has a speed $v/2$ in the positive $x$ direction. The centre of the subhalo is located in the $x$--$y$ plane at $y = b$, with its $x$ coordinate chosen to make the distance between the two centres 3~Mpc, comparable to the sum of the virial radii. The subhalo is set in motion at speed $v/2$ in the negative $x$ direction. \textbf{This is shown schematically in Fig.~\ref{fig:geometry}.}

For two halo masses $M_1$ and $M_2$, the approximate infall velocity can be analytically estimated by considering the turnaround radius $d_0$ where the relative radial velocity is 0, as shown in \citep{Sarazin2002}: 
\begin{equation}
    d_0 \simeq 4.5~\textrm{Mpc} \times \left(\frac{M_1+M_2}{10^{15}M_\odot}\right)^{1/3}\times\left(\frac{t_{\rm merge}}{10^{10}~{\rm yr}}\right)^{2/3} 
\end{equation}
\begin{equation}
    v \simeq 2930~{\rm km/s} \left(\frac{M_1+M_2}{10^{15}M_\odot}\right)^{1/2}\left(\frac{d}{1~{\rm Mpc}}\right)^{-1/2} \left[\frac{1-\frac{d}{d_0}}{1-(\frac{b}{d_0})^2}\right]^{1/2}
\end{equation}
where $d$ is the separation between the halo centres at the beginning of the simulation, $b$ is the impact parameter, \textbf{and $t_{\rm merge}$ is the age of the Universe at the time of the merger}. Since Abell 2146 is observed at z = 0.2323 and is close to pericentre passage, we use $t_{\rm merge} = 10.8$~Gyr. The infall velocity, on average, then ranges from 720 - 1220~km/s for the range of halo masses explored. 

Neither the infall velocity nor the impact parameter is directly observable post-merger, and the separations and velocities they do produce are observed in projection.  Assuming that the two BCGs in the field trace the potential minima of the two clusters, \citet{White2015} constrained the merger plane to be tilted 13-19$^\circ$ from the plane of the sky. The observed shock velocity, 2200~km/s for the bow shock, is higher than the initial velocities, since the halos and the gas in them accelerate under gravity; this point was crucial in the interpretation of the Bullet Cluster, which was otherwise considered an anomaly within $\Lambda$CDM \citep{Springel2007,Lage2015}. The X-ray observations suggest a small, non-zero impact parameter \citep{Russell2010}, but cannot constrain it directly. We test a range of values from 50-700~kpc and assess how it affects the X-ray features. 

To summarise, the initial cluster velocities are along the $x$-axis, the impact parameter along the $y$-axis, and the default line-of-sight is the $z$-axis, (0, 0, 1). If the viewing direction is defined by the polar and azimuthal angles ($\theta$, $\phi$), images we see will be projections of the simulation box along the normal
\begin{equation}
    \mathbf{n} = [\sin\theta\cdot\cos\phi, \sin\theta\cdot\sin\phi, \cos\theta].
\end{equation}

Thus, if the 3D separation between the halo centres is $\mathbf{d}$ and their 3D relative velocity is $\mathbf{v}$, the observed separation and  relative line-of-sight velocity  are given by $d_{\rm  proj} = \sqrt{\mathbf{d}^2 - (\mathbf{d} \cdot \mathbf{n})^2}$ and $v_{\rm  los} = - \mathbf{v} \cdot \mathbf{n}$. 

\section{Constraining Cluster Merger Models using X-ray and Lensing Observations}
\label{sec:obs}

\begin{figure*}
\centering
\includegraphics[height=0.35\textwidth]{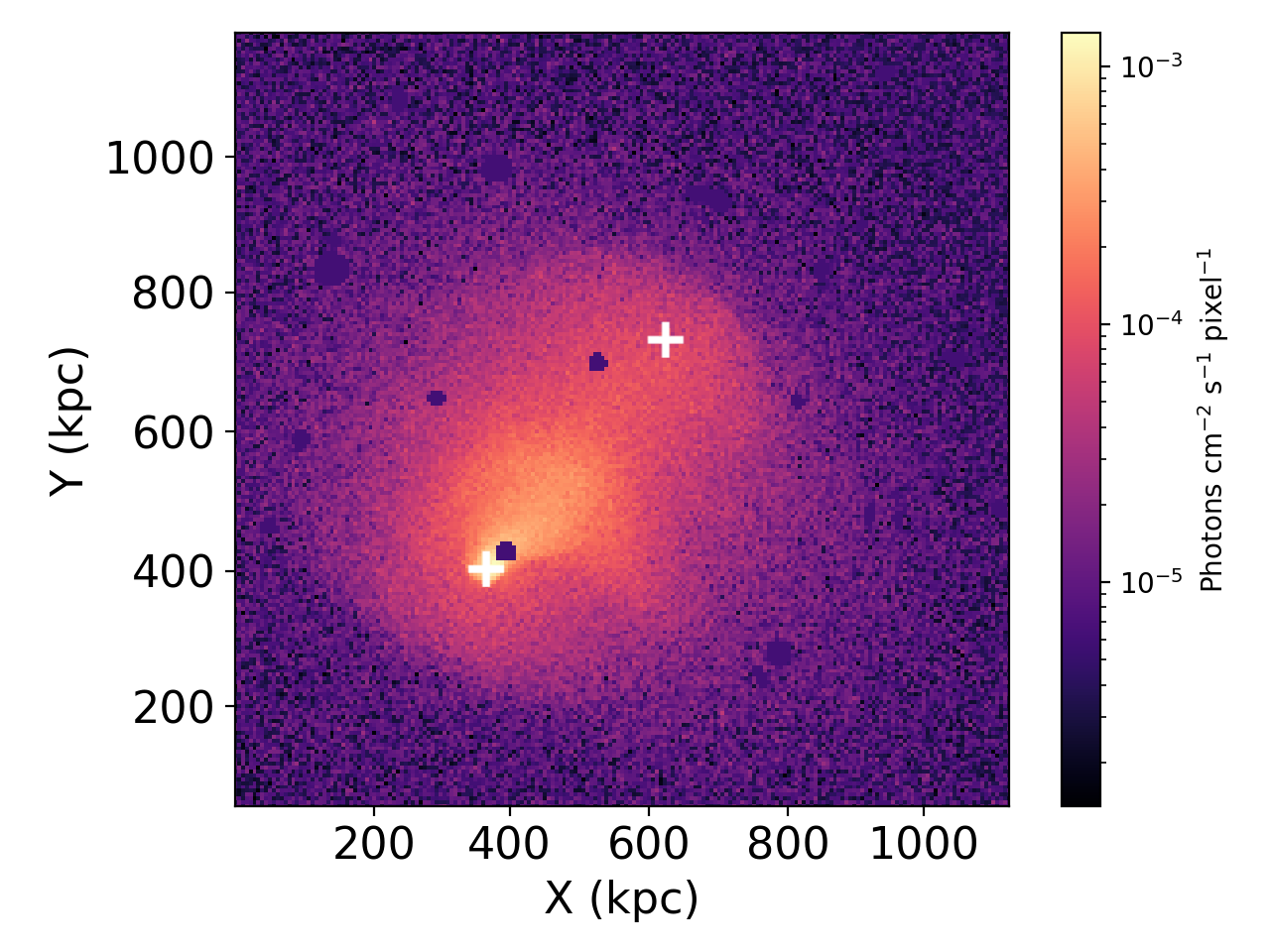}
\includegraphics[height=0.35\textwidth]{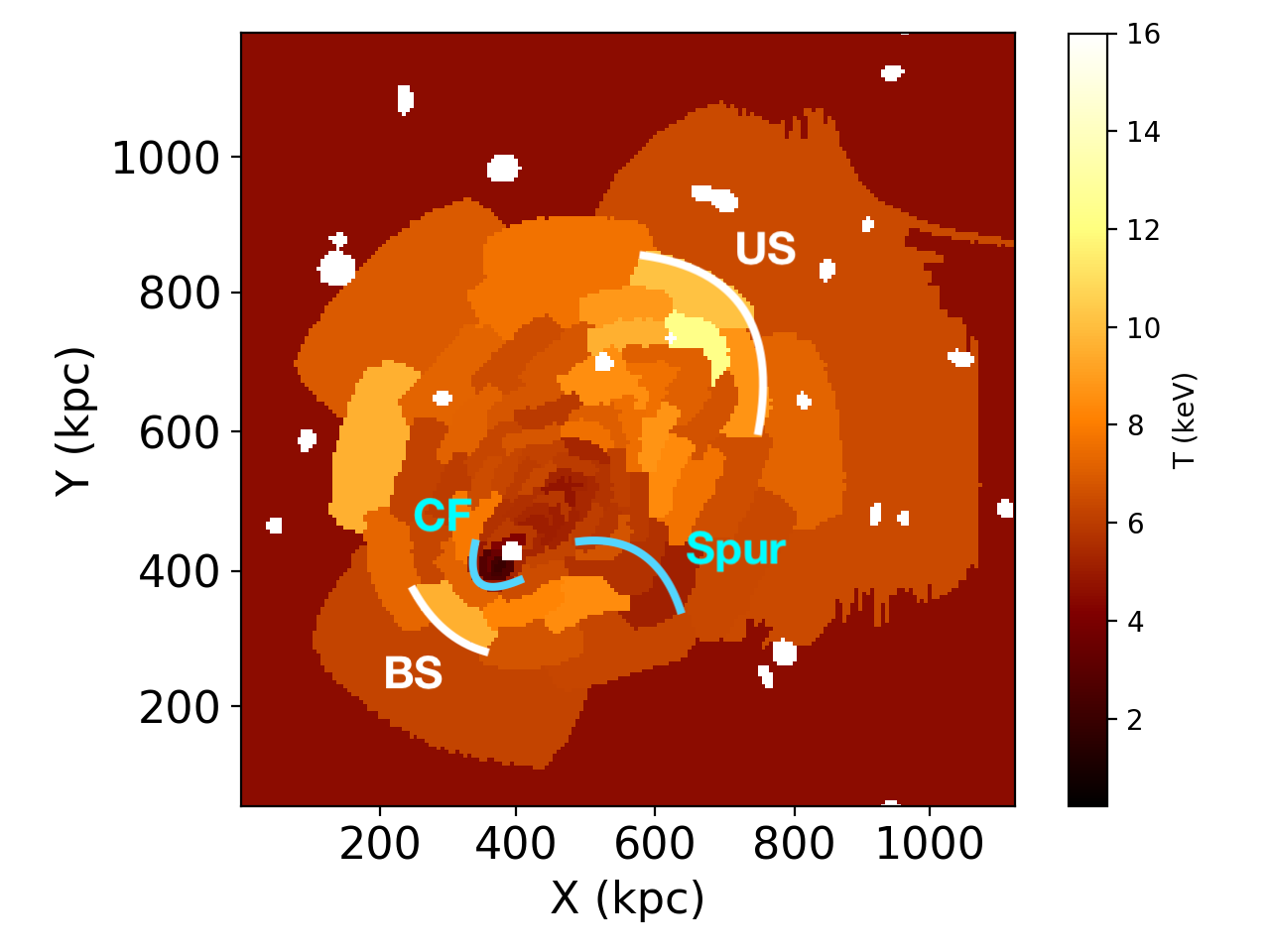}
\caption{X-ray surface brightness (left) and temperature (right) maps of Abell 2146, reproduced from the \textit{Chandra} archive. The BCGs A2146-A and A2146-B identified and used in optical and lensing studies are marked as white crosses on the surface brightness map. The bow shock and upstream shock are highlighted in white, and the cold front and plume in blue, on the temperature map. \textbf{The black areas in the surface brightness and corresponding white areas in the temperature map are where bright point sources were masked.}}
\label{fig:obs}
\end{figure*}

In this work, we focus on modeling multi-wavelength observations of Abell 2146, consisting of X-ray observations \citep{Russell2010, Russell2012}, spectroscopic data on the bright galaxies \citep{White2015}, and strong+weak lensing measurements \citep{King2016, Coleman2017}. The details of each observation can be found in the corresponding papers. Below, we highlight the salient features of these observations that are relevant to the merger and structural parameters of A2146. The quantitative constraints are summarised in Table \ref{tab:parameters}.

\subsection{The \textit{Chandra} data}
\label{sec:chandra}

We use the 419.5~ks of ACIS observations of A2146 from the {\it Chandra} archive. The data were reduced using the \texttt{chav} software package\footnote{\url{http://hea-www.harvard.edu/\textasciitilde alexey/CHAV/}} with \texttt{CIAO v4.6} \citep{Fruscione2006}, following the processing described in \cite{Vikhlinin2005} and applying the \texttt{CALDB v4.7.6} calibration files. The X-ray surface brightness map has a resolution of 1", corresponding to 3.7~kpc at the cluster redshift z$_c$ = 0.2323. The temperature maps are coarser, since it is necessary to bin several pixels together to get enough photons to fit a spectrum. Therefore, we use the surface brightness maps to identify sharp features in the ICM, and then infer from the temperature maps whether they are shock or cold fronts.

\subsubsection{Shock fronts}
A shock is an abrupt increase in both temperature and density occurring when a perturber - in this case, the infalling subcluster - moves through an ambient medium, creating and propagating a compressive disturbance. Unlike in an adiabatially compressed region, the entropy in a shock is increased.
Abell 2146 features two of these - a bow shock with a Mach number ${\cal M} = 2.3 \pm 0.2$, and an upstream shock of ${\cal M} = 2.1 \pm 0.1$ \citep[][Russell et al. in prep.]{Russell2012}. The two have slightly different formation channels, and reproducing both simultaneously is a strong constraint on our simulations. As elaborated in \S\ref{sec:evolution}, the merger produces other shocks too, although these are too weak to be detected by \textbf{current X-ray telescopes}.

Bow shocks form in front of the perturber, with the leading edge perpendicular to its direction of motion. The upstream shock is a reverse shock that forms ahead of the centre of the primary cluster. As the subcluster falls in, its halo gas is impeded by the ICM of the primary cluster and slows down. Gas stripped from around the leading edge of the merging subcluster forms an obstacle to the gas falling in its wake. The resulting pressure disturbance travelling into the wake then develops into a shock propagating upstream into the wake. We note that this shock is distinct from the leading shock that propagates away from the initial contact discontinuity through the subcluster.

The bow shock forms first, followed shortly by the upstream shock. For each shock, the Rankine-Hugoniot jump conditions yield the Mach number, which along with the distance travelled provides an estimate of the age of the shock. \citet{Russell2012} used this method to place the system at 0.1-0.3~Gyr post pericentre passage. The relation between the shock strength and age, however, is non-trivial for realistic cases where the perturber is not a rigid object, but rather a diffuse cluster that gets stripped and deformed over time \citep{Zhang2019}. The distance between the two shocks evolves rapidly following pericentre passage. This is therefore our preferred indicator of dynamical stage. 

In the optically thin ICM, a shock front only appears as a sharp feature where it is nearly tangent to our lines-of-sight. What we observe is the projection of the emission per unit volume along the line-of-sight. A shock front generally curves away from the direction of propagation. Sight lines outside the shock intersect only unshocked gas. Sight lines inside the shock intersect some shocked gas, and the depth of the shocked gas increases as the line moves further inside the front. The abrupt edge is due to the depth of the sight line within the shocked gas varying with distance, $x$, behind the front as $\sqrt{x}$, for small $x$. Only the component of the gas velocity perpendicular to the shock front is affected by the shock, so that the shock is generally strongest at its leading edge and weakens towards the periphery of the curved shock front, where the front is inclined to its direction of motion. As a result, the standoff distance between the shock and the cold front marking the boundary of the remnant gas core is smallest at the leading edge of the shock and increases towards the periphery. This causes the observed standoff distance to depend on viewing direction, increasing as our lines-of-sight tilt away from the normal to the plane of the merger \citep[see][for a more detailed review of these effects]{Markevitch2007}. 

The preshock temperature for the bow shock is close to the initial temperature of the gas at the appropriate radius in the primary cluster. The potential minimum of the primary cluster, like its observed BCG, lie within the upstream shock. The pre-shock temperature of the upstream shock, on the other hand, is that of high entropy gas from the outskirts of the subcluster, which has been subject to substantial adiabatic compression as it flowed into the central region of the primary cluster.

The strength of the bow shock depends on the movement of the perturber (i.e., the substructure core) through the ICM. This depends on the initial velocity, as well as on its gravitational acceleration due to the masses of the two halos, which in turn is larger if the impact parameter is lower. For the upstream shock, the velocity difference is between the gas stripped from the leading edge of the substructure, and the rapidly infalling gas from its outskirts. 

Given the above considerations, constraints on initial velocity, masses, impact parameter and projection angles can only be inferred once the correct snapshot, i.e.,\ dynamical phase, has been identified. Since the shock separation is the most rapidly evolving observable, we use that to select suitable snapshots.

\subsubsection{Cold fronts}
\label{sec:coldfront}

Unlike shock fronts, cold fronts are discontinuities where the temperature decreases as the density increases; in merging clusters, these are usually the remnants of cool cores \citep[][]{Markevitch2007}. If the merger were perfectly head on, the two cores would strongly disrupt, so that any cool core remnants feature would be very spread out. Furthermore, the two disrupting cores would be elongated along the same axis, that of the merger.Abell 2146 has an obvious cold front associated with BCG-A, a bullet-like subcluster punching through the ICM of the larger primary cluster, and a second, less striking discontinuity perpendicular to the axis between the shocks. These two are highlighted with blue curves in the temperature map of Fig.~\ref{fig:obs}. The second feature has been called the "plume" in the X-ray observations, which also suggested that this is most likely the disrupted gas core of the primary halo \citep{Russell2012}. Such a configuration of cold features requires a non-zero impact parameter. If the impact parameter is too large, the substructure remnant will curve significantly towards the merger axis and towards the primary core, but never pass through any part of it directly. The two cool cores would thus be left relatively intact. Thus we can constrain the impact parameter using the brightness, width, and relative orientation of the two cold features.
\begin{table}
\caption{Summary of observed constraints for Abell 2146.} 
\label{tab:parameters}
  \begin{center}
    \begin{tabular}{c|c} %
      M$_{\rm  tot, lens}^\dagger$ &1.01 - 1.36 $\times10^{15}M_\odot$\\ 
      $d_{\rm  BS-US}^{*,1}$ & 700 kpc \\
      d$_{\rm  BS-CF}^{*,2}$ & 140 kpc \\
      $(T_f,T_i)_{\rm  BS}^{*,3}$ & (10, 5) kpc\\
      $(T_f,T_i)_{\rm  US}^{4}$ & (12, 6) kpc\\
      $\Delta v_{rel}^\ddagger$ & 763 km s$^{-1}$ \\ 
      \end{tabular}
  \end{center}
 $^\dagger$ \citet{King2016} \\
 $^*$ \citet{Russell2012}\\
 $^\ddagger$ \citet{White2015}\\
 $^1$ Shock separation, i.e., distance between the points of maximum curvature of the bow and upstream shocks.\\
 $^2$ Standoff distance, i.e., distance between points of maximum curvature of the bow shock and cold front.\\
 $^3$ Pre- and post-shock temperature for the bow shock.\\
 $^4$ Pre- and post-shock temperature for the upstream shock, from Russell et al. in prep.
\end{table}

\subsection{The origin of observable merger features}
\label{sec:evolution}
\begin{figure*}
    \centering
    \includegraphics[width=\textwidth]{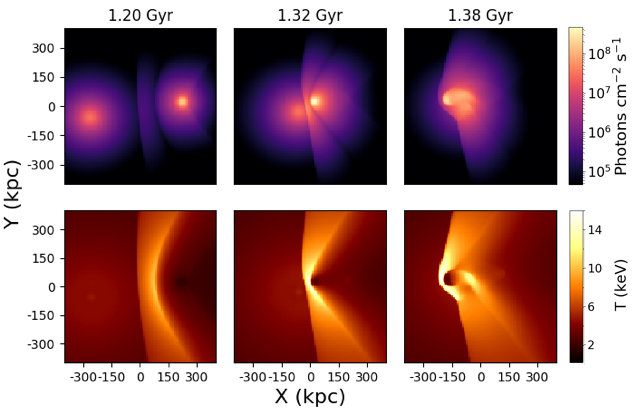}
    \caption{Snapshots of the surface brightness (top) and Mazzotta-weighted projected temperature (bottom) maps around the pericentre passage of the subcluster through the primary cluster, illustrating the development of the observed X-ray features. The first panel shows the initial contact discontinuity (i.e., a cold front), surrounded by one weak shock traveling towards the left and another towards the right. Pericentre passage occurs just before the middle panel, and the substructure core here creates an additional bow shock and contact discontinuity. By the final panel, subcluster gas that had been swept upstream by ram pressure meets gas that is still infalling, creating the upstream shock. The remnant of the primary cluster core is ejected almost perpendicular to the axis between the bow and upstream shocks, and is called the "plume" in observations.}
    \label{fig:evolution}
\end{figure*}

Fig.~\ref{fig:evolution} shows snapshots illustrating the development of the observed features. These have been described in detail in \citet{Roettiger1996, Roettiger1997,Takizawa2005, Poole2006}. Here, we present a brief summary to develop physical intuitions into the effects of the explored parameters. Illustrative snapshots are shown in Fig.~\ref{fig:evolution}. In the first panel, we see that as the subcluster falls in from the right, high entropy-gas from the outskirts of the two clusters is compressed into a high-temperature region around the contact discontinuity. This is enveloped by an extended, weakly shocked region. The centre of the subcluster, and therefore its BCG, originally lies outside this shocked region, but accelerates toward the shocked region as the rightward shock moves towards it, so that it eventually enters the shocked region. The subhalo core overtakes and passes through the initial contact discontinuity, and drives a shock behind the leftward moving shock.\textbf{ This is seen as the yellow-white, hottest region in the middle temperature panel.} As the subcluster core undergoes pericentre passage, these two shocks connect, creating the appearance of prominent bulge near the centre of the large-scale front. This bulge is the feature identified as the bow shock in the observations. The pre-shock temperature ahead of this bow shock is that of the ICM of the primary halo. There is a second contact discontinuity between the cool core of the subcluster and the shocked ICM of the primary cluster, which is the cool core of the subcluster being elongated by ram pressure; this is what \citet{Poole2006} call a ‘comet-like tail’, and is seen in both the Bullet Cluster \citep{Markevitch2002} and Abell 2146. Some gas  stripped from the remnant subcluster core obstructs higher velocity gas falling to the left, in the wake of the subhalo, leading to the formation of the upstream shock, seen clearly in the third panel. The pre-shock gas here is from the outskirts of the subcluster, so that it has a relatively high entropy and adiabatic compression heats it well above its initial temperature. The core of the primary cluster is disrupted; this low entropy gas gets ejected perpendicular to the cold front from the subcluster core, forming the feature called a plume by \citet{Poole2006}.

Eventually, the subcluster core turns around, whereas the shocks continue to move outwards. In our simulations, as in \citet{Poole2006}, this happens $\sim$1~Gyr after the first pericentre passage. However, the shocks at this point are too weak and extended to be comparable to systems like Abell 2146. Therefore, we focus on what can be learned from mergers in the first 0.5~Gyr after pericentre passage, while their morphology resembles that of Abell 2146 and the Bullet Cluster.

We treat the brightest cluster galaxies (BCGs) as tracers of the gravitational potential minima of the two merging clusters. The optical observations, used to determine positions and velocities of the BCGs, have \textit{Hubble}'s diffraction-limited resolution of 0.05" or $\sim$0.18~kpc. Spectroscopic data is available for 63 of the brightest galaxies in the clusters \citep{White2015}, which yields line of sight velocities with uncertainties of less than 1$\%$. We note that BCGs do not generally lie exactly at the potential minima of their host clusters, especially not in the midst of a merger. The relation between the BCG and cluster core velocities, therefore, is significantly less precise.

There is a clear BCG in the bullet-like cold front, referred to in the literature as BCG-A and treated as the BCG of the primary cluster referred to as Abell 2146-A \citep[e.g.,][]{Canning2012, White2015, King2016,Coleman2017}. These studies have assumed that the second brightest galaxy in the field, shown in the left panel of Fig.~\ref{fig:obs} near the upstream shock, is the BCG of the secondary cluster; this has been called BCG-B. \textbf{Instead, our simulations consistently show that BCG-A must be associated with the secondary cluster, and BCG-B with the primary cluster, in order to reproduce the X-ray features.}

\subsection{Mass Profiles from Lensing Data}
\label{sec:lens}

\textbf{Weak lensing maps offer complementary constraints on the total mass and mass profiles of Abell 2146. These rely on measurements of the distortion of shapes of background galaxies in deep imaging with the \textit{Hubble Space Telescope (HST)} \citep{King2016}. The observed ellipticity of a background galaxy ($\epsilon$) depends on the intrinsic ellipticity ($\epsilon_{i}$), as well as the (complex) reduced shear from the the cluster lens ($g$):}
\begin{equation}
    \epsilon = \epsilon_i + g = \epsilon_i + \frac{\gamma}{1-\kappa}\,,
\end{equation}
\textbf{where $\gamma$ and $\kappa$ are the (complex) cluster shear ($\gamma \equiv \gamma_{1} + i\gamma_{2})$ and convergence, respectively. The convergence is given by $\kappa = \Sigma/\Sigma_{cr}$,}
\textbf{where $\Sigma$ is the surface mass density of the lens, and $\Sigma_{cr}$ is the critical surface mass density}
\begin{equation}
\Sigma_{cr} = \frac{c_{v}^2D_s}{4\pi G D_{ds}D_d}\,,
\end{equation}
\textbf{which depends on the angular diameter distances to the source $D_s$, to the lens $D_d$, and between the two, $D_{ds}$. The speed of light is denoted by $c_v$. $\kappa$ and $\gamma$ depend on the second derivatives of the lensing potential $\psi$:}
\begin{align}
    \kappa &= \frac{1}{2}\left(\psi_{,11} + \psi_{,22}\right)
\end{align}
and
\begin{align}
    \gamma_1 &= \frac{1}{2}\left(\psi_{,11} - \psi_{,22}\right)\\
    \gamma_2 &= \psi_{,12}.
\end{align}
In the weak lensing regime, $\kappa \ll 1$ and $|\gamma| \ll 1$. 

A strong lensing analysis of the multiple images shows that the mass peaks at the brightest galaxy in the bullet-like cold front, BCG-A \citep{Coleman2017}. The parametric models considered in the weak lensing analysis had NFW components centred on the BCGs A and B, simultaneously fit to the lensing reduced shear data (ellipticities of the distant galaxies) \citep[][]{King2016}. The free parameters were the two cluster radii $r_{\rm  200}$ or, equivalently, masses enclosed inside $r_{\rm  200}$, $M_{\rm  200}$. The field-of-view of the data ($\sim$750~kpc at the system redshift) is not sufficient to simultaneously fit the cluster concentrations and masses. \textbf{Therefore, the concentrations of the two clusters were fixed for each parametric fit, and were set to be equal to one another and in the range expected from the cluster mass-concentration relationship \citep[][]{Duffy2008}.} For $c=4$, the total virial mass of the clusters is $M_{\rm 200}=1.2\times 10^{15}M_\odot$, and their mass ratio is $\approx 4$. The total mass is about 10$\%$ bigger for $c=3.5$ and 10$\%$ smaller for $c=4.5$. In the weak lensing analysis, Abell 2146-A, centred on BCG-A, is the more massive cluster. \textbf{Instead, our simulations suggest the opposite to be the case - the deeper potential minimum, associated with the primary cluster, resides in the upstream shock, like the observed BCG-B.}

In order to fit projected mass density profiles to the simulation data, or obtain synthetic shear maps, we must project the total 3D mass in the simulation boxes along the line of sight to obtain $\Sigma$ or $\kappa$. 
Assuming that the gravitating mass profile of each cluster is well described by the NFW model \citep{Navarro1996}, we can then simultaneously fit projected NFW profiles \citep{Lokas2001} centred on the two potential minima to the projected mass using 
\begin{align}
    \Sigma(R) &= \frac{c^2}{2\pi}g(c)\frac{M_v}{r_v^2}\cdot \frac{1 - \mid c^2\tilde{R}^2 - 1 \mid ^{-1/2} C^{-1}\left[\frac{1}{c\tilde{R}}\right]}{c^2\tilde{R}^2-1},
\end{align}
where $c = c_{\rm NFW}$, $\tilde{R} = R/r_v$, $R$ is the projected distance from the centre and 
\begin{align*}
g(c) &= \frac{1}{\ln(1+c) - \frac{c}{1+c}}, \\
C^{-1}(x) &= \begin{cases}
    \cos^{-1}(x), & \text{if $R > r_s$}\\
    \cosh^{-1}(x), & \text{if $R < r_s$}.
    \end{cases}
\end{align*}

\citet{Roediger2012} have shown that the observed mass of a subcluster within a given radius can vary dramatically over the course of a merger (though they only examined a single simulation with a large impact parameter). The gravitational potential deepens, and includes the mass of both systems near pericentre, so that during pericentre passage the measured concentration would be substantially larger than pre-merger. When the system eventually virialises again with the combined mass of the two halos, the total concentration is lower than pre-merger, consistent with the concentration-mass relation. 

\textbf{There is a degeneracy between mass and concentration in weak lensing observations, such that assuming a lower concentration requires a higher mass to produce the same convergence $\kappa$. Furthermore, lensing observations do not sample the full density field - rather, they provide local measurements of the reduced shear $\gamma/(1-\kappa)$ at the positions of background galaxies. For Abell 2146, \citet{King2016} had $\sim$1500 such shape measurements. The limited number of galaxies per radial bin combines with the shape noise, i.e. inherent dispersion in the unlensed distribution of shapes for the background galaxies, to restrict the signal-to-noise ratio. Lastly, as noted above, the field of view of the HST observations did not extend to the outskirts of the cluster system. As a result, it was unfeasible to simultaneously fit the dark matter concentration and total mass/virial radius of each cluster. Therefore, the lensing analysis focused on a range of concentrations $3.5 \le c_{\rm NFW} \le 4.5$, motivated by the concentration-mass relation \citep[e.g.,][]{Duffy2008}, and then fit the virial radius for each $c_{\rm NFW}$. These masses are 50-100$\%$ higher than those from X-ray and SZ observations.}

\begin{figure}
    \centering
    \includegraphics[width=0.45\textwidth]{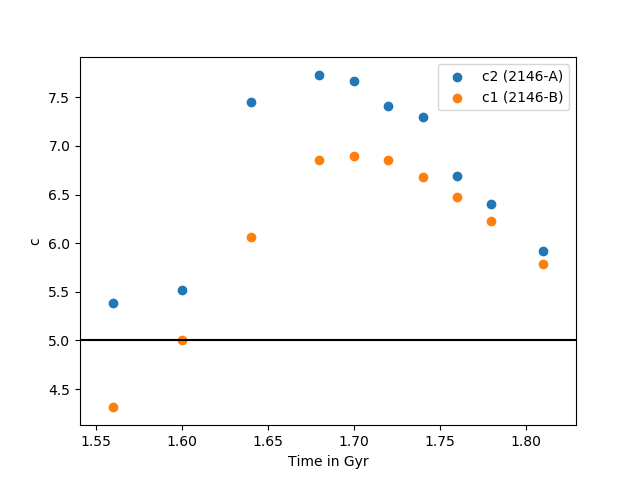}
    \caption{The evolution with time of the dark matter concentration parameter for NFW haloes fit to the simulated projected (dark matter) density. In this case, the clusters have initial true total masses $M_1 = 5\times 10^{14}M_\odot$ and $M_2=1.6\times10^{14}M_\odot$ and concentrations $c_1 = c_2 = 5$ respectively as indicated by the horizontal line.}
    \label{fig:lensing_test1}
\end{figure}

\textbf{The concentrations of merging halos, however, are systematically different from those of their relaxed counterparts. We found that initial concentrations of $3 < c_{\rm NFW} < 6$ for either halo, a range broader than the 1-$\sigma$ scatter in the $c-M$ relationships at cluster masses, was consistent with the X-ray observations. We then created maps of the projected density at 8 snapshots around the one that best fit the X-ray observations, capturing the 0.2~Gyr centred on core passage. At each snapshot, the projected density was sampled at $\approx$ 60,000 points and projected NFW density profiles were simultaneously fit to this sampling. The results of these fits are shown in the top panel of Fig.~\ref{fig:lensing_test1}, where the initial values were $c_1 = c_2 = 5$ (corresponding to clusters referred to as Abell 2146-B and Abell 2146-A respectively in the lensing papers). Assuming that a spherically symmetric NFW profile is a good description for the haloes even this close to core passage, the best-fit $c_{\rm NFW}$ is biased high, particularly for the less massive subcluster. Consistent with the results of \citet{Roediger2012}, $c_{\rm NFW}$ peaks at pericentre passage and then decreases over time. Even 0.1~Gyr after pericentre passage, the fit concentrations are $20\%$ higher than the initial values. }

\begin{figure}
    \centering
    \includegraphics[width=0.45\textwidth]{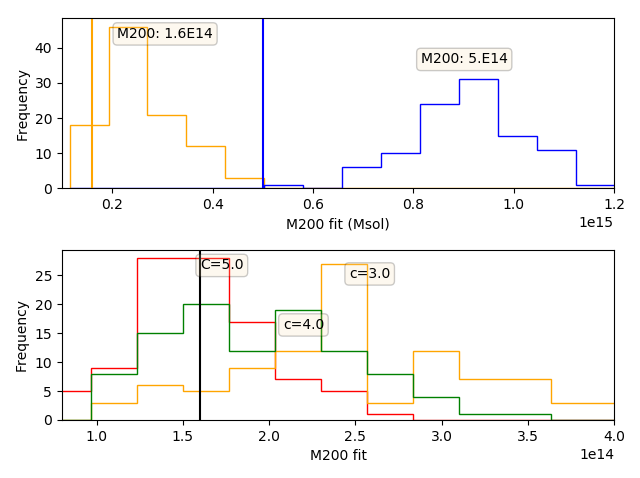}
    \caption{
    Upper panel: The distribution of masses for NFW haloes fit to synthetic lensing catalogues, generated using haloes with true masses $M_1 = 5.0\times 10^{14}M_\odot$ and $M_2=1.6\times10^{14}M_\odot$ and concentrations $c_1 = c_2= 5$. Fit distributions are shown when $c=3$ is assumed for each halo. Results for the lower (higher) mass halo are shown in orange (blue), with a vertical line indicating the true mass. The masses are overestimated by $\sim 50\%$ for the less massive halo, and $\sim 85\%$ for the more massive halo.
    Lower panel: The distribution of masses for NFW haloes fit to synthetic lensing catalogues, generated using a halo with $M_2 = 1.6\times10^{14}M_\odot$, $c_2=5$. Fit distributions are shown when $c=3$ (orange), $c=4$ (green) and $c_2=5$ (red) are assumed. The true halo mass is indicated with a black vertical line.}
    \label{fig:lensing_test2}
\end{figure}

\textbf{To illustrate and quantify the mass error due to the assumption of particular concentration values when fitting parameterised lens models, 
we created shear maps using a Fast Fourier Transform (FFT), and lensed synthetic background galaxy populations with galaxy number density set to match the observed $HST$ field. As was done in the lensing papers, we assume different values for $c_{\rm NFW}$ and fit only for the virial mass $M_{200}$.  The upper panel of Fig.~\ref{fig:lensing_test2} shows the results for haloes of mass $M_1 = 5.0\times 10^{14}M_\odot$ and $M_2 = 1.6\times10^{14}M_\odot$ and concentrations $c_1 = c_2 = 5$. Fit distributions are shown when $c=3$ is assumed for each halo. Results for the lower (higher) mass halo are shown in orange (blue), with a vertical line indicating the true mass. To demonstrate the impact of adopting a particular concentration, the bottom panel shows the results of such a lensing forward model for a halo mass $M = 1.6\times10^{14}M_\odot$ and $c = 5.0$, i.e., the best fit initial NFW parameters for the less massive subcluster. For 100 different realizations of background galaxy positions and ellipticities, the orange, green and red curves show the distributions of the fit mass $M_{200}$ assuming $c=3$, $c=4$ and $c = 5$ while fitting respectively.}

\textbf{Note that assuming concentrations lower than the true value biases mass high. During model fitting, as noted above, \citet{King2016} focused on concentrations in the range $(3.5 - 4.5)$ for the clusters in the system, and correspondingly obtained a higher mass, $(1.0 - 1.3)\times10^{15}M_\odot$ than would have been obtained when adopting higher concentrations, which is consistent with our analysis. However, as noted in \citet{King2016}, in order for the lensing data to yield a mass for Abell 2146-A similar to that of Abell 2146-B, Abell 2146-A would have to have $c_1\sim 9$ when $c_2 \sim 3.5$ for Abell 2146-B.}

\textbf{A key lesson from this exercise is that lensing masses, especially from parametric reconstruction, are degenerate with the assumed concentrations. Ideally, there would be sufficient signal-to-noise and field-of-view in the lensing data to constrain both simultaneously. But this is very rare in space-based lensing observations. Instead, the robust, model-independent quantity from lensing surveys is the reduced shear measured from the distant galaxy ellipticities. Comparisons between cluster mass model parameters from simulations and from observations must therefore be made by forward modeling the simulations and fitting their shear maps to the same parametric mass models, under the same observational conditions, such as field-of-view, galaxy number density available for shear measurements etc. }

\textbf{The simulated analogs to observable quantities are summarised in Table \ref{tab:sims}}.

\begin{table}
    \centering
    \begin{tabular}{p{0.28\linewidth} | p{0.65\linewidth}}
        \textbf{Observable} & \textbf{Simulated Analogue} \\
        Surface brightness & Emission-weighted projected photon emissivity in 0.3-7 keV\\
        Temperature & Mazzotta-weighted projected temperature \\
        Lensing & Projected density map \\
        Galaxy spectra & Average velocity of dark matter particles in 50~kpc radius (BCG) or 1 Mpc (cluster average)
    \end{tabular}
    \caption{Summary of the simulated analogues to observed quantities}
    \label{tab:sims}
\end{table}

\section{Results}
\label{sec:results}
The primary goal of this work is to investigate how the observed properties of A2146 depend on the parameters of the simulation. To this end, we compare simulations where all parameters are held constant except the one in consideration, and choose snapshots where the shock separation meets this observed constraint. For all sections but that on the viewing direction in \S\ref{sec:viewing}, the system is viewed along the $z$ axis, perpendicular to the plane of the orbit.

\subsection{Initial Cluster Merger Setup}
\begin{figure*}
    \centering
    \includegraphics[width=\textwidth]{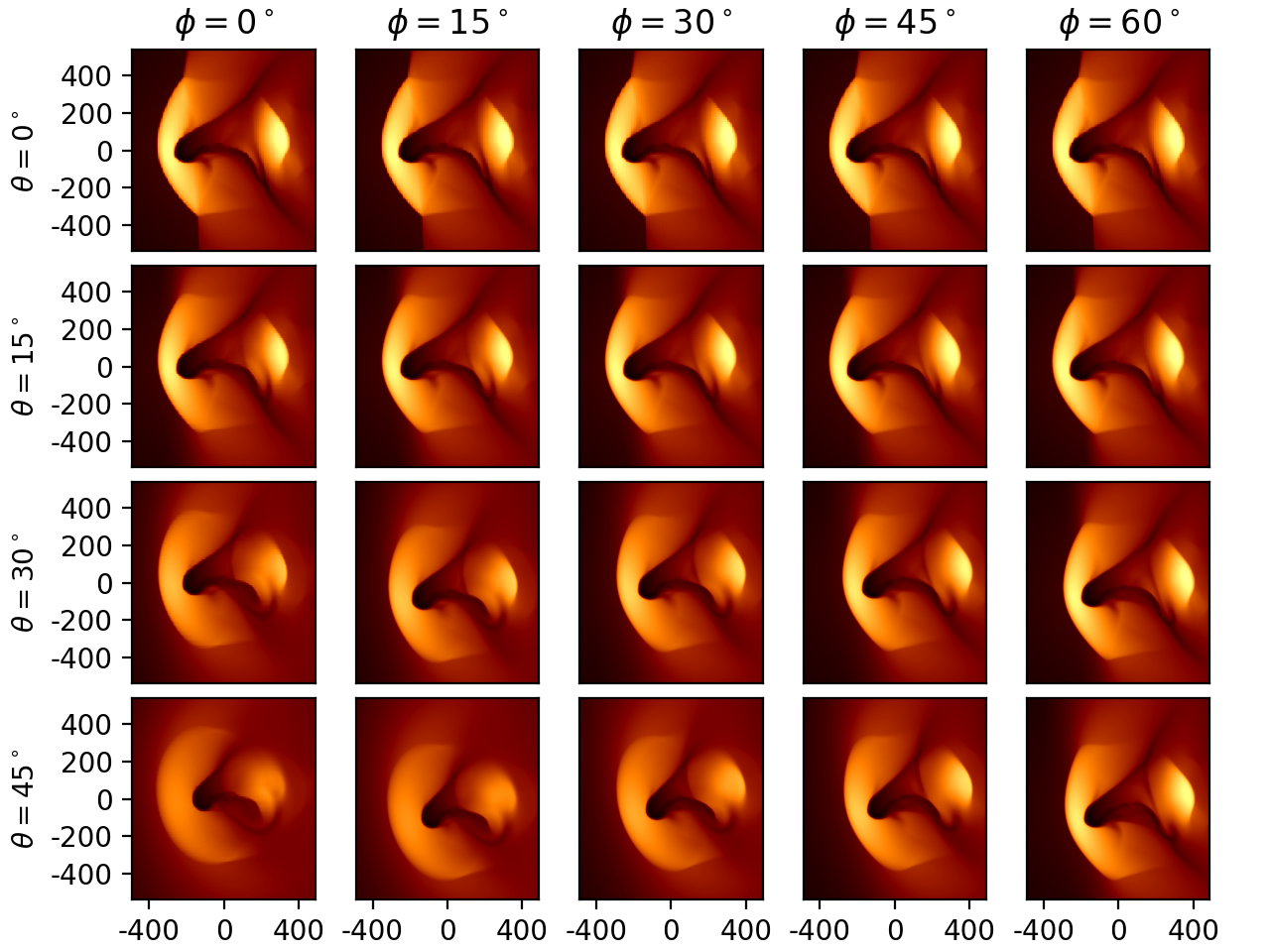}
    \caption{The effect of changing the viewing angles ($\theta$, $\phi$). Mazzotta-weighted projected temperature maps \textbf{for the simulation with $M_1 = 5\times10^{14} \Msun, ~R = 1:3, ~b=100~{\rm kpc}, ~v_{\rm rel}=1452~{\rm km/s}$}. Any viewing angle that is not perpendicular to the plane of the merger will cause weaker apparent shocks. Each panel is 800~kpc a side, and the colormap is the same as for all simulated temperature maps. We find that shocks are visible from every viewing direction shown here, i.e., visibility of shocks is not a guarantee of a nearly plane-of-sky merger. }
    \label{fig:rotation}
\end{figure*}
\textbf{Since the mass ratio of this merger derived from weak lensing has a large uncertainty due to the restricted field-of-view and the necessity to assume concentration parameters, we constrained this parameter by examining simulations from the Galaxy Cluster Merger Catalog \citep{ZuHone2018}.\footnote{http://gcmc.hub.yt} This allows us to inspect mock observations of quantities such as projected X-ray surface brightness, spectral-like weighted temperature following \citet{Mazzotta2004} (henceforth Mazzotta-weighted), and total mass density for a range of mass ratios and impact parameters in binary merger simulations. Using the simulation set ``A Parameter Space Exploration of Galaxy Cluster Mergers'' in the Galaxy Cluster Merger Catalog\footnote{http://gcmc.hub.yt/fiducial/index.html} \citep{ZuHone2011}, we identified a region of parameter space to explore further to find an analog for Abell 2146. If the subhalo is very close in mass to the primary halo, the system would look more symmetric; if the mass of the subhalo is too small, the cold front would be much weaker and the core of the primary halo \textbf{is} barely disrupted. If the impact parameter is close to zero, both cores are extremely disrupted and the standoff distance between the bow shock and cold front is too big; if it is too large, the cold remnant core of the subhalo appears extremely curved as the cores accelerate towards each other. Lastly, for the cold front and both shocks to be prominent, the observation must have occurred shortly after first pericentre passage. The X-ray features of Abell 2146 were qualitatively similar to the mergers with a mass ratio of 1:3 and a small, non-zero impact parameter, seen (0.3-0.5)~Gyr post pericentre passage. In this work, we will explore more finely around this position in parameter space, and additionally study the effects of varying dark matter and gas profiles of the halos, relative velocity, and viewing direction. }

If the BCGs are relatively good tracers of the potential minima, the orientation of the merger is well-constrained by the angles ($\theta$, $\phi$) for which the 3D separation and relative velocities of the potential minima in the simulation match observations after projection. The projected maps of surface brightness, temperature and mass should be made for the appropriate viewing direction to compare with observations. 

In addition to the BCGs, we can also use the separation between the bow and upstream shocks, which evolves rapidly, to identify a small number of snapshots for further inspection.
As shown in Fig.~\ref{fig:rotation}, we find that the shocks appear too weak if $\theta \gtrsim 30^\circ$ and $\phi\gtrsim 30^\circ$. The observed separation between the leading edges of the two shocks is 440~kpc \citep{Russell2010}. Therefore, we only keep snapshots where $440 < \textrm{d}  < 508$~kpc, where the upper bound ensures that $d \cos 30^\circ \le 440$~kpc.
\begin{figure*}
    \centering
    \includegraphics[width=\textwidth]{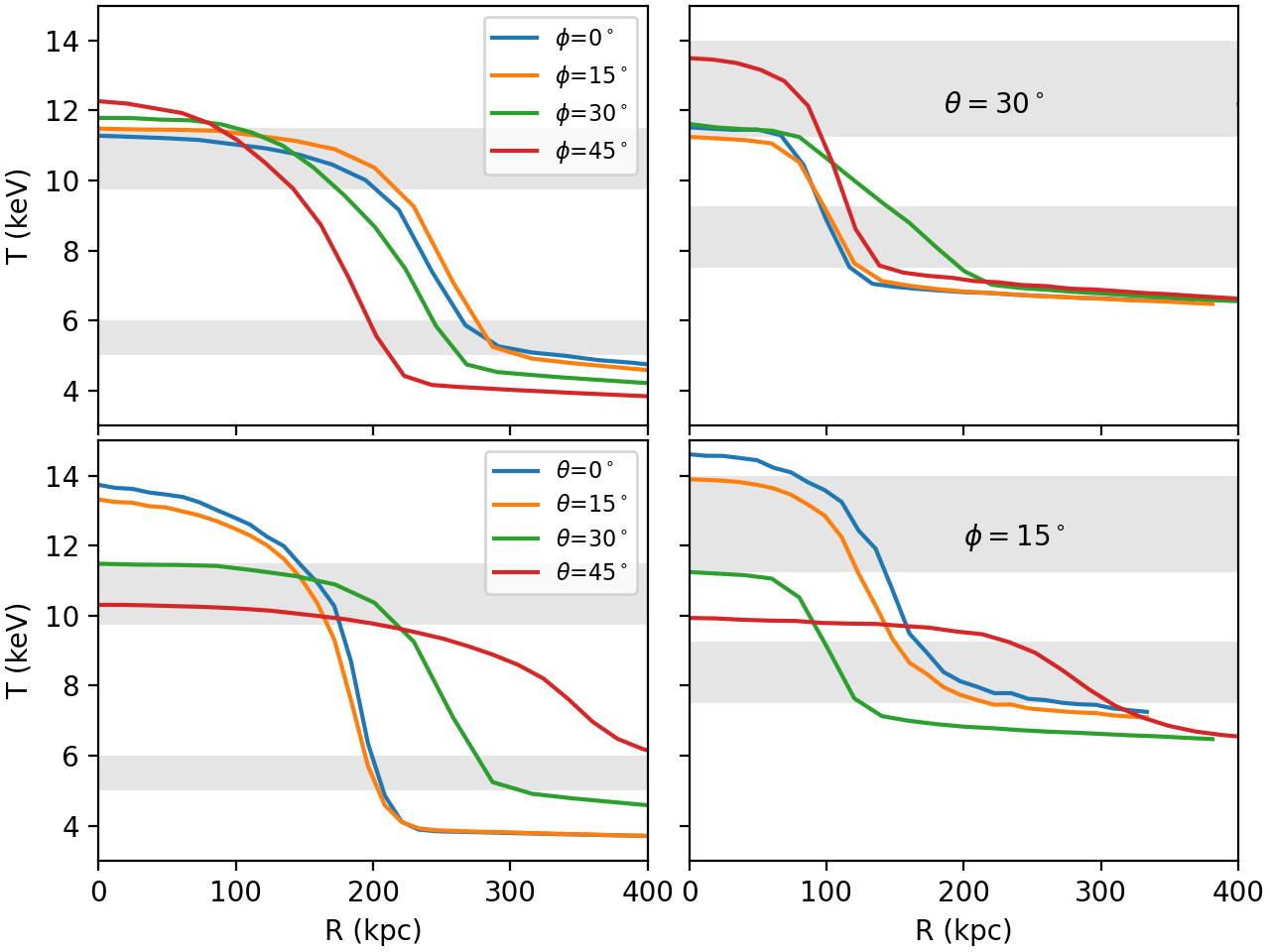}
    \caption{The temperature profiles of the bow (left) and upstream (right) shocks for different viewing directions. The top panels hold $\theta=30^\circ$ and vary $\phi$, while the bottom panels hold $\phi=15^\circ$ and vary $\theta$. The shaded grey regions show the 1-$\sigma$ error bars on the pre- and post-shock temperatures for the corresponding shocks from \citet{Russell2012}.}
    \label{fig:rotation_profiles}
\end{figure*}

The purpose of this study is to investigate how observed properties of the system depend on each of the input parameter of the simulation. Through a preliminary round of parameter tests, we chose to zoom in on the region around a primary cluster mass $M_1 = 6\times10^{14}M_{\odot}$, a mass ratio $R = 1:3$, an impact parameter $b = 100$~kpc, and initial relative velocity $v_{rel} = 1500$~km/s. \textbf{The following sections describe the zoomed in region of parameter space, so as to focus on systems like Abell 2146.}

\subsection{Viewing direction}
\label{sec:viewing}

As shown in Fig.~\ref{fig:geometry}, the merger in our simulations occurs in the x-y plane of the simulation domain, with the initial relative velocities along the x-axis and initial impact parameter along the y-axis; this defines the so-called plane of the orbit. The default line-of-sight is in the z-direction, as is the angular momentum. Instead, if the observer views the system along some different direction, they would see a different projection of the 3D system. Each viewing direction is defined by viewing angles $(\theta, \phi$), where $\theta$ is the polar angle and $\phi$ the azimuthal angle (see Fig.~\ref{fig:geometry}).

Fig.~\ref{fig:rotation} illustrates the effect on the appearance of a simulation of changing the two viewing angles \textbf{for a given snapshot. For small values of $\phi$, increasing $\theta$ (a) increases the apparent pre-shock temperature, (b) increases the stand-off distance between the cold front and the bow shock and (c) decreases the apparent offset between shocks. The latter effect is the smallest, because the two shocks have large radii of curvature}. Increasing $\phi$ has a barely discernible effect for low $\theta$, but as seen in the panels for $\theta \gtrsim 30^\circ$, counters the effect of changing $\theta$ alone. Each of these effects can be explained by simple geometric arguments. 

\begin{figure}
    \centering
    \includegraphics[width=0.45\textwidth]{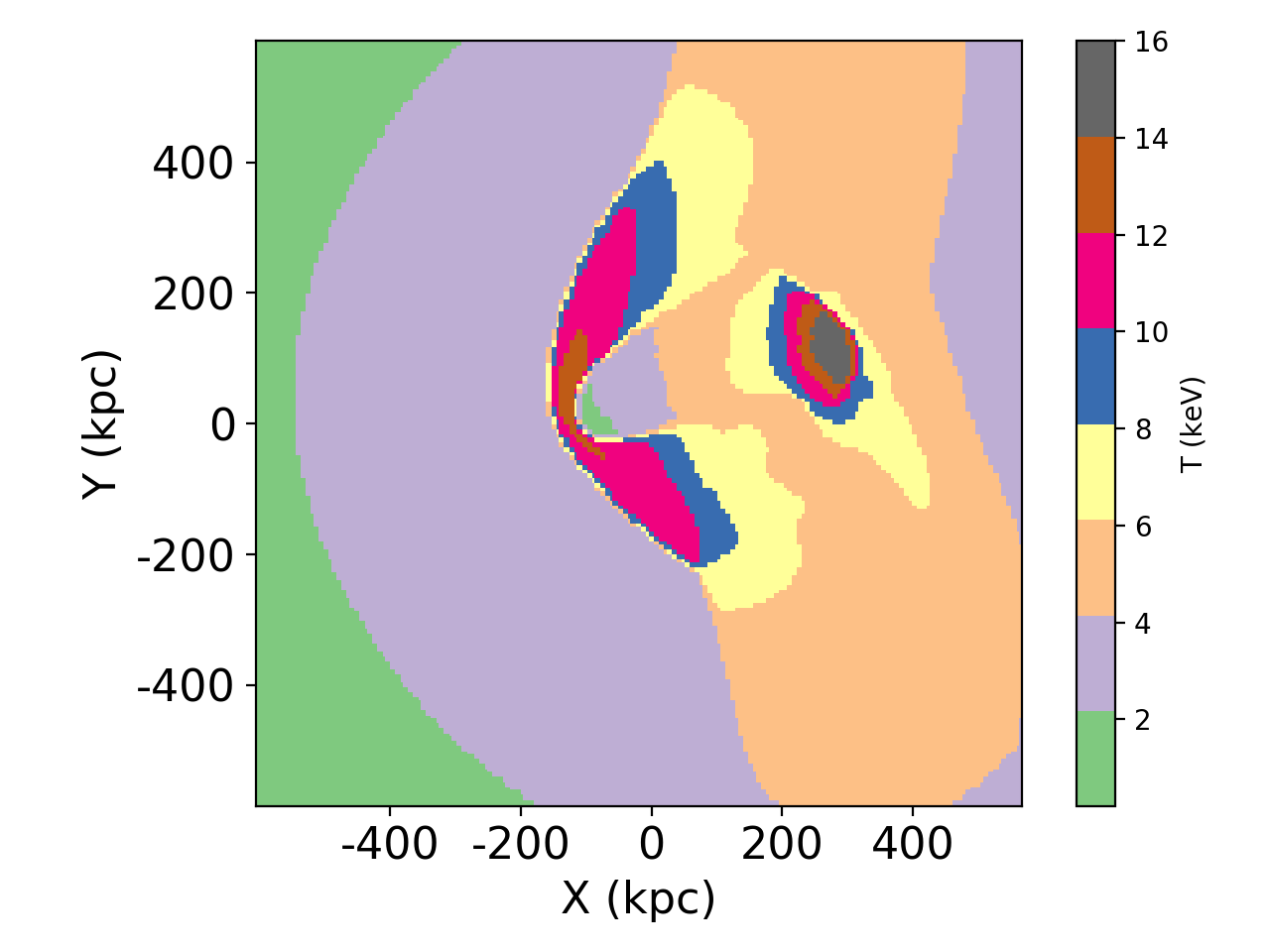}
    \caption{Temperature map for one of the snapshots similar to Abell 2146, using a discrete colormap to emphasise the gradient in pre-shock temperature. Changing the viewing angle would lead us to see the shock at a smaller distance from the primary cluster core, where the background temperature is higher. As a result, the shock will appear weaker.}
    \label{fig:isophotes}
\end{figure}

The shocks form along the axis of motion of the substructure through the ICM, in the x-y plane, so their separation is maximal along that axis. The velocities of the two subclusters are also entirely in the x-y plane, so that if viewed along the z-axis, the line-of-sight velocity difference between the subclusters is 0. The unperturbed cluster is spherical, and the radius of curvature $R_s$ of the shock front is smaller than its cluster-centric radial position $r$. The unshocked gas can be assumed to move at approximately the same speed around the shock front, but the jump conditions only apply to the component of the gas velocity perpendicular to the shock front at any point. The speed of the shock at each point on the front is therefore $v \cos\eta$, where $\eta$ is the angle between the normal to the front and the velocity of the front with respect to the gas; thus $\eta = 0^\circ$ at the leading edge of the shock. Changing the viewing direction generally moves the point where our line of sight is tangent to the shock front away from the leading edge of the shock, so that the observed shock strength is weaker. This additionally moves the tangent point to smaller $r$, increasing the preshock temperature at the tangent point. Finally, the radius of curvature $R_c$ of the cold front is smaller than $R_s$. This means that the minimum separation between the two features, i.e.,the standoff distance, is minimised in the plane of the merger, and increases for other viewing directions. Each of these phenomena can be understood intuitively as illustrated in Fig.~\ref{fig:isophotes}, which shows a simulated temperature map with a discrete colormap to accentuate the difference in temperature depending on the position $r$.

Older simulation studies, tailored to the observing capabilities of telescopes like ROSAT, stated that the merger needs to be close to the plane of the sky for the shock features to be visible \citep[e.g.,][]{Ensslin1998, Ricker1998}, although how close is not clearly defined. Fig.~\ref{fig:rotation} shows, however, that both shock fronts are distinctly visible even for inclinations as high as $(\theta = 30^\circ, \phi = 15^\circ$). Therefore, the visibility of shocks does not constrain the merger axis to be very close to the plane of sky. 

Instead, we need a more quantitative comparison, as shown in Fig.~\ref{fig:rotation_profiles}. Note that this is for $M_1 = 5\times10^{14}M_\odot, M_2 = 1.6\times10^{14}M_\odot$. Higher masses would increase the normalisation of the temperature profiles. \textbf{The horizontal shaded areas show the 1-$\sigma$ error bars on the pre- and post-shock temperatures for Abell 2146 (Russell et al, in prep.). From Fig.~\ref{fig:rotation}, we find that $\theta=30^\circ$ produces the correct standoff distance at $\phi=0$. The top panel of Fig.~\ref{fig:rotation_profiles} shows the temperature profiles across the bow (left) and upstream (right) shocks for various values of $\phi$ at $\phi=30^\circ$, with good matches to the observations for $\phi < 15^\circ$. The bottom panel then holds $\phi=15^\circ$ and varies $\phi$. In this way, we find that $\theta = 30^\circ$  matches observations. Lower values result in a post-shock temperature that is too high, while higher values cause the shock to be much shallower than observed.}

The standoff distance between the bow shock and the cold front also depends on the relative velocity, or, equivalently, the angular momentum of the merger. \textbf{To break this degeneracy, we aim to additionally reproduce the observed line-of-sight velocity offset between the BCGs \citep{White2015}. Our simulations do not explicitly include cluster galaxies; however, BCGs are known to trace the potential minima of galaxy clusters \citep[e.g.,][and references therein]{Zitrin2012}. The potential minima of the halos were identified using the \texttt{peak-local-max} function in the Scikit-Image Python package on the slice of the gravitational potential in the x-y plane. The velocity $v$ of a BCG is estimated as the average velocity of all the dark matter particles within 50~kpc of its potential minimum; this radius is characteristic for BCGs of clusters of the masses considered \citep[e.g.,][and references therein]{Lin2004}. The line-of-sight velocity difference between the BCGs depends on the viewing angle, as detailed in \S \ref{sec:geom}.}

\subsection{Total mass and mass ratio}
\label{sec:masstests}

\begin{figure*}
    \centering
    \includegraphics[width=\textwidth]{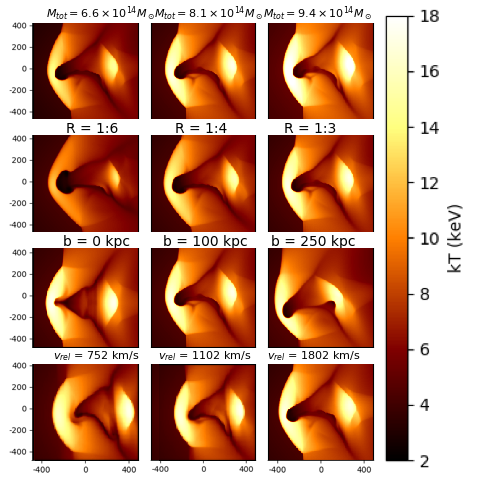}
    \caption{The effect of changing key dynamical parameters one at a time, holding all the others constant. \textit{Top row:} Increasing the total mass of the system increases the temperature of the ICM as well as the shocks. \textit{Second: }Increasing the mass of the secondary cluster, and therefore the mass ratio $R=M_1:M_2$, increases the strength of the shocks as well as increasing the standoff distance. \textit{Third: }Increasing the impact parameter $b$ (in kpc), while holding all other parameters constant, curves the path of the secondary cluster core towards the primary core, creating a more arced cold front. The relative sizes and orientations of the different X-ray features are most significantly affected by $b$. \textit{Bottom row:}  Increasing the initial relative speed of the cluster centres increases the strengths of the shocks, but reduces the standoff distance and changes the radii of curvature of the shock fronts.}
    \label{fig:paramtests}
\end{figure*}

Increasing the total mass, first of all, increases the overall projected temperature, since the thermal pressure now has to balance a greater weight of overlying gas. The Mazzotta-weighted average temperature of A2146 within a radius of 2 arcmin ($\simeq 440$~kpc), covering both the shocks, is 7.5 $\pm$ 0.3 keV if the cool core (of radius 10" or 37~kpc) is excised and 6.7 $\pm$ 0.3 keV if it is included \citep{Russell2012}. Simply by matching the limits of the colorbars in the observed and simulated maps of projected temperature, we can visually rule out systems whose average temperature is too small or too large. The core-excised (included) average temperatures for the 1:3 mass ratio mergers presented in Fig.~\ref{fig:paramtests} are 5.62 (5.15) for $M_{\rm tot} = 6.6\times10^{14}M_\odot$, 7.33 (6.23) for $M_{\rm tot} = 8.1\times10^{14}M_\odot$ and 8.10 (6.30) for $M_{\rm tot} = 9.4\times10^{14}M_\odot$, where $M_{\rm tot} = M_1 + M_2$ is the sum of the two total masses in the super-NFW formulation. The corresponding total virial masses are $(4.5, 5.4, 6.3)\times10^{14}M_\odot$, with the middle value consistent with observed temperatures assuming that the merger occurs in the plane of the sky. Fig.~\ref{fig:paramtests} further reminds us that if the merger does not occur in the plane of the sky, the observed temperature is an underestimate. Therefore, a greater mass is possible if the viewing angles are larger.

\begin{figure}
    \centering
    \includegraphics[width=0.45\textwidth]{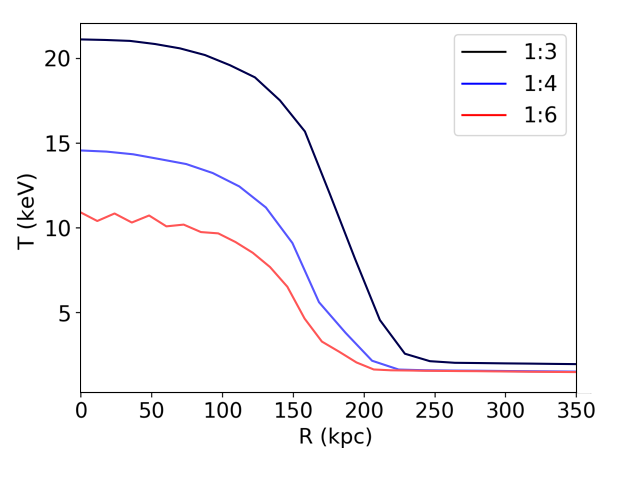}
    \caption{Projected temperature profiles of the bow shock for same primary halo mass but different mass ratios. Not only is the peak temperature significantly higher for a more massive secondary halo, but the shocked region is wider.} 
    \label{fig:massratio_profiles}
\end{figure}
Secondly, increasing the total mass increases the scale radii of the two halos. This means that for the same absolute magnitude of the impact parameter, greater fractions of the two cluster cores interact with each other during pericentre passage. In other words, increasing halo mass while holding impact parameter constant is equivalent to holding mass constant and reducing impact parameter, so that the cores are more disrupted. In the second row of Fig.~\ref{fig:paramtests}, this is seen as a "fatter" bullet and a less prominent plume, either due to higher total mass at fixed mass ratio (top row), or higher mass ratio at fixed primary halo mass (second row). The acceleration due to increased halo mass also means that for the same initial relative speed in the simulation, the shocks produced at the best-fit snapshot are stronger. Fig.~\ref{fig:massratio_profiles} additionally quantifies the difference between using different mass ratios but same mass for the primary halo, by plotting the temperature profiles out from the peak of the bow shock. 

Increasing the mass of the secondary also increases ram pressure stripping of the secondary core. This can be seen in narrowing of the leading edge of the core, the formation of a stronger upstream shock, and a larger standoff distance. Based on the second row of Fig.~\ref{fig:paramtests}, then, we favour a mass ratio of 1:3. The standoff distance is still lower than in observations, but, as discussed above, this can be fixed with a larger $\theta$.

\subsection{Impact parameter and initial relative speed}
The X-ray observations rule out a perfectly head-on merger because the disrupted subcluster core is asymmetric. The head of the bullet is curved, and its tail fans out more towards the South than to the North. The greater the impact parameter, the greater this asymmetry. The appearance of this bullet is thus affected by the orientation of our line of sight with respect to the plane of the orbit. Fig.~\ref{fig:rotation} showed this for the case $b=100$~kpc. If viewed from sufficiently close to the plane of the orbit, the curvature of the bullet becomes very hard to perceive, and it becomes difficult to distinguish from a merger with a zero impact parameter. For larger $b$, however, the curvature of the bullet is too large to be erased by modest inclination of the merger plane with the plane of the sky. Furthermore, the core of the subcluster experiences very little ram pressure, and the cold front is much wider than observed. Similarly, the core of the primary cluster is less perturbed for larger $b$, leaving an intact core rather than a "plume". Given these effects, we can constrain $b \sim 100$~kpc.

As seen in the last panel of Fig.~\ref{fig:paramtests}, increasing the relative velocity of the perturber increases the strength of the shock. The effect on the standoff distance is less linear. On the one hand, if the subcluster moves faster through the ICM, it stays closer to the bow shock, and this decreases the standoff distance decreases. We see this effect as we increase $v_{rel}$ from 720 to 1252 km/s. But increasing the velocity also increases the ram pressure, pushing gas from the subcluster core away from its direction of motion and increasing the standoff distance. This is what we see in further increasing $v_{rel}$ to 2200 km/s. Given the observed strength and width of the upstream shock, the intermediate speed of 1252 km/s is most likely; the observed standoff distance can then be increased by increasing $\theta$, as is already encouraged by the analyses of total mass and mass ratio.

\begin{figure*}
    \centering
    \includegraphics[width=0.85\textwidth]{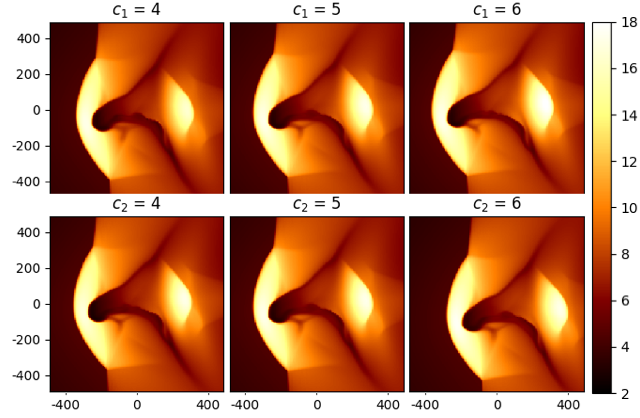}
    \caption{The effect of changing the dark matter concentration of the primary (top) and secondary (bottom) halo while holding all other parameters constant. A lower concentration for the primary, or a higher one for the secondary, results in more instabilities along the cold strip of gas connecting the two disrupted cores. We note that the details of KHI are sensitive to the presence of turbulence in the ICM, which we have not included. All other features remain unaffected.}
    \label{fig:c1}
\end{figure*}

\subsection{Dark matter concentration}
\begin{figure*}
    \centering
    \includegraphics[width=\textwidth]{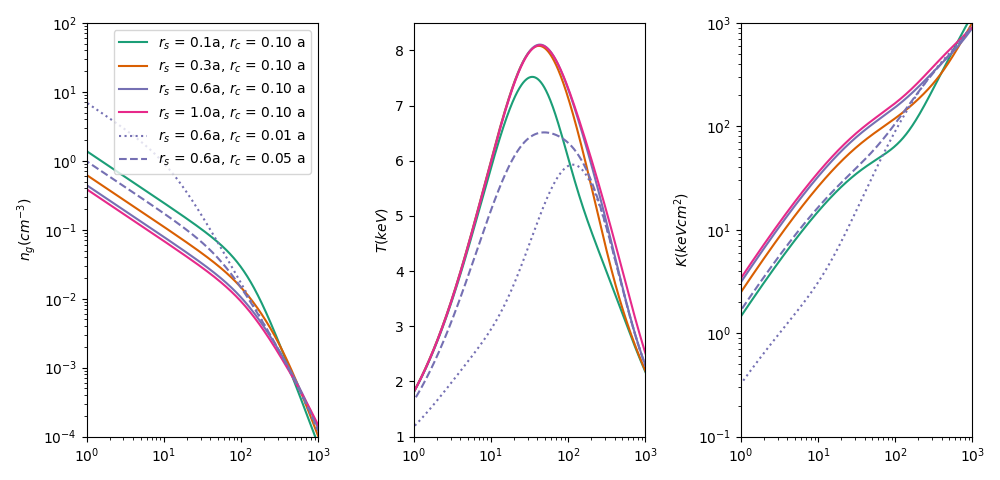}
    \caption{The effect of varying the gas scale radius $r_s$ in Eq.~\ref{eq:vikhlinin} and core radius $r_c$ for  $\alpha=2$. Here, $M_{sNFW} = 1.6\times 10^{14}M_\odot$, $c_{\rm NFW}= 5.2$, and $r_c = 0.1a$. Decreasing either the scale or core radius produces a denser, lower-entropy core. While there is a spread in observed cluster cores, we find that the cool core population is generally well fit with $\alpha=2, r_c \sim 0.05a$ and $r_s = 0.6a$, so that the entropy decreases monotonically towards the centre. Through a similar comparison, we describe non-cool cores with $\alpha=0$, $r_c\sim0.3a$ and $r_s=0.6a$.}
    \label{fig:ic_paramtest}
\end{figure*}

A higher concentration for either cluster makes its core more compact and resistant to disruption. If the primary cluster is more concentrated, there is more gravitating mass enclosed within the core, and the subcluster bullet is accelerated more during infall. As a result, the gas in the subcluster experiences higher ram pressure $p_{\rm ram} = \rho v^2$. This causes more gas to be swept away from the leading edge of the subcluster core, into a wake, which is undergoing a reverse shock. This has two observable consequences. Slowing the gas "bullet" increases the standoff distance between the bow shock and the cold front. At the same time, the gas displaced from the core of the infalling cluster impedes the gas infalling from further in its wake, boosting the strength and extent of the upstream shock. Both of these effects are shown in the top panel of Fig.~\ref{fig:c1}. The bottom panel shows that the plume
feature associated with the core of the primary halo is brighter and less disturbed if its concentration is higher. We find that the concentration of the subcluster, on the other hand, has no appreciable effect on the gas observables. Since other parameters affect the same observables much more dramatically, we find that $c_2$ is not well-constrained by X-ray imaging.

\begin{figure*}
    \centering
    \includegraphics{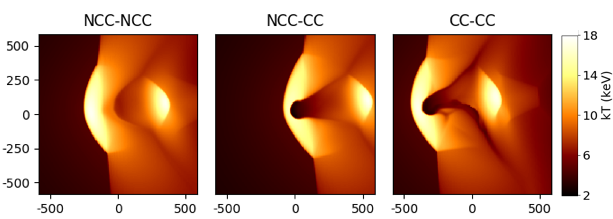}
    \caption{The morphology of the merger looks extremely different depending on whether each of the components has a cool or non-cool core. Here, both clusters are modeled with $\alpha=0$; the cool cores have core radii $r_c = 0.02a$, and the non-cool cores have $r_c=0.5a$, where $a$ is the NFW scale radius. Abell 2146 clearly resembles a system where both components were initially cool-core clusters.}
    \label{fig:cc-ncc-test}
\end{figure*}

We note that this result is for mergers of the explored mass ratios, 1:3-1:6. If a merger is closer to equal mass, $c_2$ would have much the same effect as $c_1$ on the gas in the other merging component. 

The standoff distance was smaller than observed for the parameters tested in Fig.~\ref{fig:paramtests}, which all used $c_1 = 4.1$. A denser dark matter halo, like $c_1 = 5$, would solve this issue and remove the need for larger $\theta$. 

\subsection{Gas profiles}
\label{sec:gasprof}
The model in Eq.~\ref{eq:vikhlinin} contains two parameters that affect the compactness and cuspiness of the core - the core radius $r_c$ and the central density slope $\alpha$. The scale radius $r_s$ and the outer slope parameters $\beta$ and $\epsilon$ affect the outskirts, so we do not vary them in our study and just use the best-fit values from \citet{Vikhlinin2006}.

It is important to note that $r_c$ is a purely empirical parameter, which can be arbitrarily tuned in the \citet{Vikhlinin2006} formulation to match the data. We would like to choose core radii that produce profiles analogous to observed relaxed clusters \citep{DeGrandi2002, Vikhlinin2005, Hogan2017}. As shown in Fig.~\ref{fig:ic_paramtest}, our profiles look like cool-cores, with high central densities and temperatures decreasing in the centre, for $\alpha = 0,~r_c/a$ = 0.02; for $\alpha = 0,~r_c/a$ = 0.5, they have lower, flat central densities and high central temperatures, like observed non-cool cores. While we acknowledge that the cores of clusters do not follow a strict dichotomy, we use these pairs of parameters when modelling each halo as a cool or non-cool core.

Fig.~\ref{fig:cc-ncc-test} shows that the X-ray images of the merger depend strongly on whether, per our modeling above, each cluster has a cool- or non-cool core. In the top panels, the primary halo has a cool core, whereas in the bottom panels it has a non-cool core. Similarly, in the left panels the subcluster has a cool core, while on the right it has a non-cool core. The four scenarios are strikingly different. When either core is non-cool, it is more extended and more susceptible to stripping. A cool-core secondary cluster will produce a bullet-like cold front; there is no low-entropy gas to form such a feature if it were a non-cool core. The remnant of the primary cluster core remains partially intact if it starts out as a cool-core, with the remainder drawn out into a plume-like shape if the secondary also has a cool core, as seen in Abell 2146. If it is a non-cool core, there is no low-entropy material to start with and form the plume. We therefore conclude that both the primary and secondary clusters in Abell 2146 must have had cool cores before the merger. Small adjustments of $r_c$ around the best-fit value of $0.02a$ then have very small effects on the width of the bullet (subcluster core remnant) and brightness and extent of the plume (primary cool remnant).

\subsection{Best fit simulations for Abell 2146}
\label{sec:bestfit}
\textbf{Our simulations are consistent with the X-ray observations for a primary halo of virial mass $M_1 = 5.0\times10^{14}M_\odot$ and an infalling halo mass $M_{2} = 1.6\times10^{14}M_\odot$, so that the mass ratio R = 1:3, observed 0.1~Gyr after pericenter passage}. Both clusters initially have cool cores. The larger mass in each case would require the merger to be inclined with respect to the plane of the sky. The concentration of the more massive halo is $\sim$ 5, on the higher end of the scatter in the concentration-mass relation; the concentration of the subhalo does not visibly affect either the X-ray or lensing maps. The initial relative speed of the cluster centres was likely  $v_{\rm rel} \simeq$ 1200~km/s, and impact parameter $b$ = 100~kpc. The system is likely viewed from a direction of $\theta = 30^\circ$; $\phi = 15^\circ$ then reproduces the observed temperature profiles at the shock as well as the line-of-sight velocity offset between the BCGs. 

\section{Discussion}
\subsection{Error bars including covariance}
Although, in principle, an error region could be constructed for the model of A2146, this is not feasible with current computing resources. Since the parameters affect the same features in different ways, mapping out the interdependence of observed features on the parameters would require a large suite of simulations sampling many combinations of all the significant parameters, including M$_{1}$, R, b, v, $\theta$, $\phi$, $c_1$. This leaves us with a 7-dimensional space even after fixing the parameters $c_2$, $\alpha_1$, and $\alpha_2$, which which have less visible impact on the X-ray images. For each simulation in such a study, it will be important to quantify the similarity of each snapshot to the observation. This would have to involve some combination of at least the shock separation, the standoff distance, the shock strengths, and the average temperature with and without the cool core. Even exploring just three values for each simulation yields over 2000 simulations, which cannot be inspected manually in the same way as in this pilot study. Instead, it would require a pipeline to compare simulations to observations and move in the parameter space. Nevertheless, here we have explored the physical impact of each of these parameters on a binary merger like A2146, demonstrating the observable impacts of each parameter on the system and providing a basis for the interpretation of similar systems in the future.

\begin{figure*}
    \centering
    \includegraphics{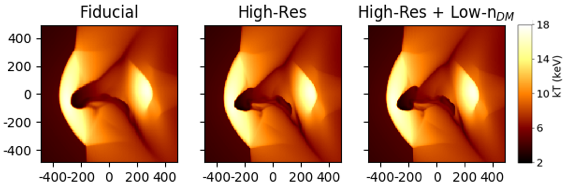}
    \caption{The effect of increasing the maximum refinement level (middle panel) and then halving the number of dark matter particles (right panel), in comparison with the resolution of this study (left panel). Doubling the hydrodynamic resolution increases the growth of fluid instabilities all along the cold discontinuities. However, the shock separation, standoff distance, average temperature, shock Mach numbers and the velocity offset between the BCGs remain unchanged. Reducing the force resolution by halving the number of dark matter particles has a negligible effect on the KHI.}
    \label{fig:restest}
\end{figure*}

\textbf{\subsection{Resolution tests}
We tested the effect of adding an additional refinement level, i.e., doubling the spatial resolution and octupling the mass resolution. Increasing the resolution primarily reduces numerical viscosity, permitting faster growth of fluid instabilities, notably the Kelvin-Helmholtz Instability (KHI). Indeed, Fig.~\ref{fig:restest} shows that the boundaries of the bullet-like cold front are more disrupted, and clear wave-like structures appear in the bridge connecting the two cool-core remnants. This will be significant in future studies that measure the plasma viscosity based on the development of KHI. It will also affect measurements of thermal conduction, which rely on the width of the contact discontinuities; this width cannot, of course, cannot be lower than the simulation resolution. The distances between the features, however, remain unchanged, as do the temperature and velocity structures. Since these are the properties we use to constrain parameters in this study, we conclude that the hydrodynamic resolution of 6.8~kpc is sufficient. Reducing the number of dark matter particles by half also did not change the results; for similar studies in the future, we would recommend using this lower number of dark matter particles to reduce the simulation time by $\sim 40\%$.}
\section{Conclusions}
\label{sec:conclusions}

We have performed a large suite of idealised simulations of binary mergers between galaxy clusters using the GPU-accelerated adaptive mesh refinement code GAMER-2, with the goal of constraining merger parameters using deep Chandra X-ray and lensing observations of Abell 2146. We assess the roles of halo masses, NFW concentrations, gas profiles, impact parameter, initial relative velocity, and viewing direction on observable quantities on X-ray properties and gravitational lensing observations. In searching for a simulated analog of the observed cluster merger Abell 2146, we find various results that will be helpful for any future interpretations of merging galaxy clusters. These will allow us to understand the laboratory, before using it as a test site for constraining cosmology and ICM microphysics. Our main findings are summarized as follows:

\begin{itemize}

\item The average temperature including and excluding the cool core pointed to a virial mass of $M_{1}=5.0 \times 10^{14} \Msun$ for the primary halo and $1.6\times10^{14} \Msun$ for the secondary halo.
    
\item The large standoff distance favours an initial infall speed of $v_{\rm rel} = 1150\rm\ km\ s^{-1}$ and a viewing direction offset from the perpendicular to the plane of the merger (the $z$-axis) by about $30^\circ$ in the initial direction of motion of the infalling subcluster [$\theta = 30^\circ$]. The observed strengths of the bow and upstream shocks are then reproduced for $15^\circ < \phi < 30^\circ$. This viewing direction also brings the simulated velocity offset between the cluster potential minima in agreement with the observed line-of-sight velocity difference of the BCGs. 
    
\item If the primary cluster has a cool core, it is more efficient at stripping the secondary core, resulting in a stronger and brighter upstream shock than if it had a non-cool core. If the secondary cluster has a cool core, it is more resilient to stripping, and disrupts the core of the primary cluster to form a plume-like feature. If instead it is a non-cool core, almost all the gas is stripped into an upstream shock behind the primary core, which in turn remains almost intact. We conclude that the cold "bullet" and "plume" features are the remnant cores of the secondary and primary clusters, respectively, if both clusters initially had cool cores and fell in with an impact parameter of $b = 100$~kpc. 
    
\item Smaller subcluster masses, smaller initial relative velocities, and larger impact parameters all result in lower Mach numbers for the shocks.
    
\item In principle, increasing the dark matter concentration of the primary halo slightly strengthens the upstream shock and increases the standoff distance of the bow shock, i.e.,~its separation from the cold front. This is because the bullet is slowed by the greater ram pressure. However, using $3 < c_{\rm NFW}< 6$ for both clusters did not produce significant enough differences in the simulated X-ray images. Therefore, X-ray images alone are unable to constrain the dark matter concentrations of the halos.
    
\item We find that the total mass of A2146 is significantly lower than previous determinations based on weak lensing data, and that (driven by the X-ray measurements) the more massive cluster is Abell 2146-B. The former factor can be explained since the parameterised models assumed NFW profiles with concentrations expected for relaxed cluster halos, whereas the core of the merging system is gravitationally compressed around pericentre passage, effectively increasing the concentration of the NFW model for each halo. Instead, it is important to consider lower halo masses than suggested from the earlier lensing analysis, and forward model using the simulations to obtain the synthetic lensing signal. That Abell 2146-A is the more massive cluster in the lensing analysis is still unexplained and is beyond the scope of this work.

\end{itemize}

This study paves the way for extracting more information from X-ray and optical observations for merging galaxy clusters. We have demonstrated how X-ray measurements alone can tightly constrain the halo masses even in a non-equilibrium system. The time since pericentre passage, impact parameter, dark matter concentration of the primary halo, and viewing direction can all be constrained using X-ray maps alone, and can be corroborated with optical measurements of BCG positions and velocities. The masses and velocities of merging clusters provide tests of cosmological models, which will be crucial in ongoing and upcoming surveys like eROSITA, DES, HSC, and Rubin. 
Once these hydrodynamic parameters have been constrained, a given merging cluster can then be used to study the nature of dark matter and ICM microphysics, such as viscosity, thermal conductivity, and magnetic field strength.

\section*{Acknowledgements}
We thank the anonymous referee for their very helpful comments. We thank Rebecca Canning for her pointers on the dynamical and lensing observations of the system, and Miyoung Choi for guiding us through the lensing analysis pipelines. UC was supported as a Chandra Pre-Doctoral Fellow by NASA Grants G08-19110B and G08-19108X. JAZ and PEJN acknowledge support through Chandra Award Number G04-15088X
issued by the Chandra X-ray centre, which is operated by the Smithsonian Astrophysical Observatory for and on behalf of NASA under contract NAS8-03060. SF and LJK acknowledge support for this work provided by the National Aeronautics and Space Administration through Chandra Award Number G08-19110D issued by the Chandra X-ray Center, which is operated by the Smithsonian Astrophysical Observatory for and on behalf of the National Aeronautics Space Administration under contract NAS8-03060. The simulations were run on the Grace HPC cluster at the Yale Centre for Research Computing. 

\section*{Data Availability}

Snapshots of the simulations used for this project will sequentially be added to the Cluster Merger catalog \citep{ZuHone2018}, accessible at \url{http://gcmc.hub.yt/}. The full simulation files are too large to host permanently on a server, and will be shared on reasonable request to the corresponding author. 

\bibliographystyle{mnras}
\bibliography{reference}

\label{lastpage}
\end{document}